# Investigation of effective thermoelectric properties of composite with interfacial resistance using micromechanics-based homogenisation

Jiyoung Jung[1], Sangryun Lee[1], Byungki Ryu[2], and Seunghwa Ryu[1,*]

**Affiliations**

[1]Department of Mechanical Engineering, Korea Advanced Institute of Science and Technology (KAIST), 291 Daehak-ro, Yuseong-gu, Daejeon 34141, Republic of Korea

[2]Energy Conversion Research Center, Korea Electrotechnology Research Institute (KERI), 12 Bulmosan-ro 10beon-gil, Seongsan-gu, Changwon-si, Gyoengsangnam-do 51543, Republic of Korea

[*]Corresponding author e-mail: ryush@kaist.ac.kr

**Abstract**

We obtained the analytical expression for the effective thermoelectric properties and dimensionless figure of merit of a composite with interfacial electrical and thermal resistances using a micromechanics-based homogenisation. For the first time, we derived the Eshelby tensor for a spherical inclusion as a function of the interfacial resistances and obtained the solutions of the effective Seebeck coefficient and the electrical and thermal conductivities of a composite, which were validated against finite-element analysis (FEA). Our analytical predictions well match the effective properties obtained from FEA with an inclusion volume fraction up to 15%. Because the effective properties were derived with the assumption of a small temperature difference, we discuss a heuristic method for obtaining the effective properties in the case where a thermoelectric composite is subjected to a large temperature difference.

## Introduction

There is increasing interest in thermoelectric power generation, because approximately two-third of the energy used in the world is lost as thermal energy.[1] Thermoelectric devices can convert thermal energy directly into electrical energy and vice versa without mechanical parts or phase change and thus can be a reliable solution to recycle the discarded thermal energy.[2] In addition, the thermoelectric generator is scalable, as it can be made large by connecting several modules in series.[2] However, thermoelectric devices are not yet commercially available at a large scale, owing to their low efficiency. The efficiency of thermoelectric materials is often related to a dimensionless number called the figure of merit (ZT), which can be enhanced when the material has a large Seebeck coefficient, high electrical conductivity, and low thermal conductivity.[3]

Recently, considerable efforts have been made to fabricate thermoelectric composites possessing a variety of inclusions. For example, nanoscale inclusions are reported to selectively reduce the thermal conductivity, while the electrical conductivity and Seebeck coefficient are less affected, which enhances the ZT.[1,4] Larger-scale inclusions that are comparable to or smaller than the longest mean free path of acoustic phonons may improve the thermoelectric efficiency by reducing the thermal conductivity. Nonetheless, it is challenging to fully exploit the aforementioned advantage of the composite owing to the large fraction of the matrix-inclusion interface inducing interfacial electrical and thermal resistances.

Because of the lack of a theoretical model to predict the effective thermoelectric properties of a composite, the prediction and design of the composite has not been conducted systematically. For a proper description of such a thermoelectric composite, it is necessary to develop a homogenisation theory accounting for three aspects that have not been considered in previous studies on the effective thermoelectric properties of a composite:[5-7] first, the transport properties of inclusions smaller than or comparable to the longest phonon mean free path; second, the thermoelectric properties in the presence of interfacial electrical and thermal resistances to account for the contribution of the interfacial area in a composite; and third, the temperature-dependent thermoelectric properties, as thermoelectric devices are usually subjected to a large temperature difference between the hot and cold sides.

In this study, as the first step towards the development of an appropriate homogenisation theory, we present a micromechanics-based solution accounting for the interfacial electrical and thermal resistances (depicted in Figure 1) in the regime where the Fourier conduction law is valid, which is thus applicable when the inclusion is larger than the longest mean free path of acoustic phonons. For mathematical simplicity, the temperature-dependent thermoelectric properties are not taken into account. We formulate the governing equation of the thermoelectric composites coupling electrical and thermal responses via the Seebeck coefficient using a $6 \times 6$ thermoelectric property matrix with six-dimensional input and output fields (electric fields and temperature gradient as input; electrical current density and entropy flux as output) and derive the corresponding Eshelby tensor as a function of the interfacial electrical and thermal resistances. Via the mean-field homogenisation scheme called the Mori–Tanaka method, we derive the closed-form solutions for the effective thermoelectric properties, including the Seebeck coefficient and the electrical and thermal conductivities of the composite, which are related to the ZT. All the solutions are benchmarked against finite-element analysis (FEA), showing that the theoretical prediction matches the FEA results up to an inclusion volume fraction of 15% (<3% errors for all the considered material properties).

Although we consider the temperature-independent thermoelectric properties, the governing equation explicitly depends on the temperature. When the average temperature between the prescribed temperatures at the cold and hot boundaries is used as the constant effective temperature in the theoretical derivation, the effective thermoelectric property prediction well matches the numerical simulation performed with a large temperature difference. The robustness of such a simple approximation originates from the very small multiplicative factors in the temperature-dependence terms in the closed-form solutions of the effective thermoelectric properties. To the knowledge of the authors, this is the first study to derive the effective thermoelectric properties of a composite in the presence of interfacial resistances and can pave the way for the systematic development of thermoelectric composites with enhanced efficiency.

## Results and Discussion

### Effective Thermoelectric Properties in Absence of Interfacial Resistance

In the steady state, the constitutive equations for the electrical and thermal conduction are coupled as follows:

$$\boldsymbol{J_E} = \boldsymbol{\sigma e} + \boldsymbol{\sigma\alpha g} \tag{1}$$

$$\boldsymbol{J_S} = \boldsymbol{\alpha\sigma e} + \frac{\boldsymbol{\gamma}}{T}\boldsymbol{g}, \tag{2}$$

where $\boldsymbol{J_E}$ is the electric current density, $\boldsymbol{q}$ is the heat flux, $T$ is the temperature, $\boldsymbol{J_S}(=\boldsymbol{q}/T)$ is the entropy flux, $\boldsymbol{e}(=-\boldsymbol{\nabla}\phi)$ is the electric field, with $\phi$ being the electric potential, $\boldsymbol{g}(=-\nabla T)$ is the temperature gradient field, $\boldsymbol{\sigma}$ is the electrical conductivity, $\boldsymbol{\alpha}$ is the Seebeck coefficient, $\boldsymbol{\kappa}$ is the thermal conductivity with zero electric field, and $\boldsymbol{\gamma}(=\boldsymbol{\kappa}+T\boldsymbol{\alpha\sigma\alpha})$ is the thermal conductivity with zero electric current. $\boldsymbol{J_E}$, $\boldsymbol{J_S}$, $\boldsymbol{q}, \boldsymbol{e},$ and $\boldsymbol{g}$ are vector fields with three components.[5,6] $\boldsymbol{\sigma}, \boldsymbol{\alpha}, \boldsymbol{\kappa},$ and $\boldsymbol{\gamma}$ are 2nd-order tensors that can be represented as $3\times 3$ symmetric matrices. $\Phi$ and $T$ are scalar fields. The constitutive equations can be expressed via the dummy index notation, as follows:

$$\begin{bmatrix} \boldsymbol{J_E} \\ \boldsymbol{J_S} \end{bmatrix} = \begin{bmatrix} \boldsymbol{\sigma} & \boldsymbol{\alpha\sigma} \\ \boldsymbol{\alpha\sigma} & \boldsymbol{\gamma}/T \end{bmatrix} \begin{bmatrix} \boldsymbol{e} \\ \boldsymbol{g} \end{bmatrix} \tag{3}$$

$$J_{Pi} = E_{PiMn} Z_{Mn} \iff \boldsymbol{J} = \boldsymbol{EZ}, \tag{4}$$

$$\text{where } J_{Pi} = \begin{Bmatrix} J_i^E & P = 1 \\ J_i^S & P = 2 \end{Bmatrix} : \text{ flux vector } \boldsymbol{J}$$

$$Z_{Mn} = \begin{Bmatrix} e_n & M = 1 \\ g_n & M = 2 \end{Bmatrix} : \text{ field vertor } \boldsymbol{Z}$$

$$E_{PiMn} = \begin{Bmatrix} \sigma_{in} & P = 1, M = 1 \\ \sigma_{ik}\alpha_{kn} & P = 1, M = 2 \\ \alpha_{ik}\sigma_{kn} & P = 2, M = 1 \\ \gamma_{in}/T & P = 2, M = 2 \end{Bmatrix} : \text{ thermoelectric matrix } \boldsymbol{E}.$$

Here, $\boldsymbol{J}$ and $\boldsymbol{Z}$ are six-dimensional output and input fields, respectively, and $\boldsymbol{E}$ is a $6 \times 6$ matrix accounting for the linear relationship between the input and output fields. The lowercase indices correspond to 1–3 (representing the three components of the Cartesian coordinate system), and the uppercase indices correspond to 1 and 2 (indicating electrical and thermal quantities, respectively). The repeatedly multiplied index (dummy index) represents the summation over all the values of the index. The notation for the constitutive equation is inspired by previous studies involving micromechanical analyses of piezoelectric composites.[8,9]

We aim to obtain the effective thermoelectric property when a composite consisting of a matrix and spherical inclusions is subjected to prescribed temperature boundary conditions at two ends with $T_H$ and $T_C$, as shown in Figure 2. For the application of the micromechanics-based mean-field homogenisation, it is necessary to have a uniform $\boldsymbol{E}$ within each constituent material. Hence, following the approaches used in previous studies[5-7], we assume a very small temperature difference within the specimen and approximate $T$ as the average temperature $T_{avg}$, as follows.

$$T_H \approx T_C \approx T \approx T_{avg} \tag{5}$$

(We will later show that the less stringent approximation $T \approx T_{avg} = \frac{1}{2}(T_H + T_C)$ for significantly different $T_H$ and $T_C$ is sufficient to ensure a good match between the theoretical prediction and the numerical calculations of the effective properties.) According to the assumption, the governing equation in a homogeneous thermoelectric material can be expressed as follows:

$$-\frac{\partial}{\partial x_i} J_{Pi} + f_P = 0 \text{ where } f_P = \begin{Bmatrix} \dot{Q} & P = 1 \\ \dot{g}/T_{avg} & P = 2 \end{Bmatrix}, \tag{6}$$

where $\dot{Q}$ is the point electric current source, $\dot{g}$ is the point heat source, and $\dot{g}/T_{avg}$ is the point entropy source. Accordingly, we can define the Green's function $\boldsymbol{G}$ for the thermoelectricity, where $\boldsymbol{G}(\boldsymbol{x}, \boldsymbol{y})$ represents the potential at $\boldsymbol{x}$ caused by the source at $\boldsymbol{y}$, as follows:

$$E_{PiMn} \frac{\partial G_{MR}(\boldsymbol{x} - \boldsymbol{y})}{\partial x_i \partial x_n} + \delta_{PR}(\boldsymbol{x} - \boldsymbol{y}) = 0. \tag{7}$$

More specific interpretations are given as follows.

- $G_{11}(\boldsymbol{x}, \boldsymbol{y})$: the electric potential at $\boldsymbol{x}$ due to a unit point electric current source at $\boldsymbol{y}$
- $G_{12}(\boldsymbol{x}, \boldsymbol{y})$: the electric potential at $\boldsymbol{x}$ due to a unit point entropy source at $\boldsymbol{y}$
- $G_{21}(\boldsymbol{x}, \boldsymbol{y})$: the temperature at $\boldsymbol{x}$ due to a unit point electric current source at $\boldsymbol{y}$
- $G_{22}(\boldsymbol{x}, \boldsymbol{y})$: the temperature at $\boldsymbol{x}$ due to a unit point entropy source at $\boldsymbol{y}$

Having defined the Green’s function, we can compute the Eshelby tensor, which was originally devised in Eshelby’s seminal paper on the elasticity. When an inclusion subjected to a eigenstrain (i.e., prestrain of freestanding inclusion) is embedded in an infinite matrix with the same elastic property, the constrained strain due to the matrix is related to the eigenstrain according to the Eshelby tensor.[10,11] Analogously, the Eshelby tensor in thermoelectric problems relates the constrained field vector $\boldsymbol{Z}$ and the eigenfield $\boldsymbol{Z}^*$ as follows:

$$Z_{Rm} = S_{RmPn} Z^*_{Pn}, \tag{8}$$

$$\text{where } \boldsymbol{Z}^*(\boldsymbol{x}) = \begin{Bmatrix} Z^* & \boldsymbol{x} \subset V \\ 0 & \boldsymbol{x} \not\subset V \end{Bmatrix}.$$

$$S_{RmPn} = \frac{\partial}{\partial x_m} \int_V \frac{\partial G_{MR}(\boldsymbol{y} - \boldsymbol{x})}{\partial y_i} E_{PiMn} \, dV(\boldsymbol{y}) \tag{9}$$

Here, $V$ is the volume of the inclusion whose thermoelectric properties are identical to the matrix. The Eshelby tensor is a function of the inclusion shape and the material properties of the matrix. The eigenfield vector $\boldsymbol{Z}^*$ can be considered as a fictitious electric field and temperature gradient field produced without an external source for a freestanding isolated inclusion, and the field vector $\boldsymbol{Z}$ is the field within the inclusion when it is embedded in the matrix.

According to the mathematical analogy with the piezoelectric formulation, the Eshelby tensor for thermoelectricity for an ellipsoidal inclusion can be written as[8]

$$S_{MnAb} = \frac{1}{4\pi} E_{PiAb} \int_{-1}^{1} \int_{0}^{2\pi} \left(\frac{\xi}{a}\right)_i \left(\frac{\xi}{a}\right)_n K^{-1}_{MP} d\theta d\xi_3 \tag{10}$$

$$\text{where } K_{MP} = \left(\frac{\xi}{a}\right)_s \left(\frac{\xi}{a}\right)_t E_{MsPt}, \qquad \frac{x_1^2}{a_1^2} + \frac{x_2^2}{a_2^2} + \frac{x_3^2}{a_3^2} = 1$$

$$\xi_1 = (1 - \xi_3^2)^{1/2} \cos\theta, \; \xi_2 = (1 - \xi_3^2)^{1/2} \sin\theta, \; \xi_3 = \xi_3,$$

where $a_i$ represents the length of the half-axis of the ellipsoid inclusion. For the spherical inclusion within an isotropic matrix, the Eshelby tensor can be simplified as

$$S_{MnAb} = \frac{1}{3} \delta_{MA} \delta_{nb}, \tag{11}$$

where $\delta_{ij}$ refers to the Kroneker delta (see Supplementary Note 1 for the derivation).

The Eshelby tensor concept can be applied to solve a problem including an inhomogeneity, which refers to an inclusion having a different thermoelectric property $\boldsymbol{E_1}$ from the property of the matrix $\boldsymbol{E_0}$, by considering equivalent fields. The inhomogeneity can be replaced with the inclusion by applying the equivalent eigenfield vector $\boldsymbol{Z}^{*,\boldsymbol{eq}}$, which satisfies the following equation:

$$\boldsymbol{J} = \boldsymbol{E_1 Z} = \boldsymbol{E_0}(\boldsymbol{Z} - \boldsymbol{Z}^{*,\boldsymbol{eq}}), \tag{12}$$

where $\boldsymbol{J}$ and $\boldsymbol{Z}$ are the flux vector and field vector in an inhomogeneity, respectively.

For multiple inhomogeneities, the Mori–Tanaka method is used for predicting the effective material properties of a composite by considering the interaction between the inhomogeneities.[11] Each inhomogeneity is assumed to be embedded effectively in the matrix subjected to a spatial average field of the matrix over the entire specimen (strain for elasticity, temperature gradient in thermal conduction), which is why the technique is referred to as the mean-field homogenisation method. Because the interaction between the inhomogeneities becomes intense at a high volume fraction of inhomogeneities, the Mori–Tanaka method is known to be reliable at a relatively low volume fraction of inhomogeneities (<20%). In this mean-field homogenisation scheme, the average fields in the matrix, inhomogeneity, and composite ($\overline{\boldsymbol{Z_0}}$, $\overline{\boldsymbol{Z_1}}$, and $\overline{\boldsymbol{Z}}$, respectively) are related as follows:

$$\overline{\boldsymbol{Z_1}} = \boldsymbol{A}\overline{\boldsymbol{Z_0}} = \boldsymbol{A}(c_0\boldsymbol{I} + c_1\boldsymbol{A})^{-1}\overline{\boldsymbol{Z}}, \tag{13}$$

where $\mathbf{A} = \left[\boldsymbol{I} + \boldsymbol{S}\boldsymbol{E}_{\mathbf{0}}^{-\mathbf{1}}(\boldsymbol{E_1} - \boldsymbol{E_0})\right]^{-\mathbf{1}}$ is the concentration tensor.

The effective thermoelectric matrix $\boldsymbol{E_{eff}}$ can be obtained from the relationship between the volume-averaged input and output fields within the composite—$\overline{\boldsymbol{Z}}$ and $\overline{\boldsymbol{J}}$, respectively—as follows.

$$\boldsymbol{E_{eff}} = (c_0\boldsymbol{E}_0 + c_1\boldsymbol{E}_1\boldsymbol{A})(c_0\boldsymbol{I} + c_1\boldsymbol{A})^{-1}$$
$$= \boldsymbol{E_0} + c_1[(\boldsymbol{E_1} - \boldsymbol{E_0})^{-1} + c_0\boldsymbol{S}\boldsymbol{E_0^{-1}}]^{-1} \tag{14}$$

where $\boldsymbol{E_{eff}} = \begin{bmatrix} \boldsymbol{\sigma_{eff}} & \boldsymbol{\alpha_{eff}\sigma_{eff}} \\ \boldsymbol{\alpha_{eff}\sigma_{eff}} & \boldsymbol{\gamma_{eff}}/T_{avg} \end{bmatrix}$, $\quad c_0 + c_1 = 1$

Here, $c_0$ is the volume fraction of the matrix, and $c_1$ is the volume fraction of the inclusion (see Supplementary Note 2 for the derivation). For thermoelectric composites made of two isotropic materials, we obtain the closed-form expressions for the effective Seebeck coefficient, $\boldsymbol{\alpha_{eff}}$, and electrical and thermal conductivities, $\boldsymbol{\sigma_{eff}}$ and $\boldsymbol{\kappa_{eff}}(= \boldsymbol{\gamma_{eff}} - T_{avg}\boldsymbol{\alpha_{eff}\sigma_{eff}\alpha_{eff}})$ (see Supplementary Note 3 for the explicit expressions). When $\alpha_0 = 0, \alpha_1 = 0$, the closed-form solutions for the effective electrical and thermal conductivities are identical to the solutions obtained for a conventional composite conduction problem in the absence of the thermoelectric effect, as follows.

$$\sigma_{\mathrm{eff}} = \frac{\sigma_0(2c_0\sigma_0 + \sigma_1 + 2c_1\sigma_1\ )}{2\sigma_0 + c_1\sigma_0 + c_0\sigma_1} \tag{15}$$

$$\kappa_{\mathrm{eff}} = \frac{\kappa_0(2c_0\kappa_0 + \kappa_1 + 2c_1\kappa_1\ )}{2\kappa_0 + c_1\kappa_0 + c_0\kappa_1} \tag{16}$$

The theoretical prediction for ZT is calculated using the following definition.

$$(\mathrm{ZT})_{\mathrm{eff}} = \frac{\sigma_{eff}\alpha_{eff}^2}{\kappa_{eff}}T_{avg} \tag{17}$$

We conduct FEA to validate the effective thermoelectric property solutions derived above. Using the Digimat software, 10 different representative volume elements (RVEs) are generated for volume fractions of 5%, 10%, and 15%, as shown in Figure 3. The length of one side of the matrix is set to 10 mm, and the diameter of the inclusion is set to 1 mm for the RVE. For the volume fractions of 5%, 10%, and 15%, 95, 186, and 286 spherical inclusions are used, respectively. The effective material properties for each RVE are obtained from the COMSOL program, and the mean values and error bars of the effective electrical and thermal conductivities and the effective Seebeck coefficient are presented in Figure 4. In the FEA, the average temperature is set to 300 K, and the temperature difference between the hot side and the cold side is set to 1% of the average temperature (3 K) for consistency with our assumption. To test the validity of the equation using an arbitrary composite having very different thermoelectric properties, we design the matrix to have the 300 K properties of $p$-type ($\alpha > 0$) bismuth telluride ($Bi_2Te_3$) with a peak ZT value of 0.397, which is a widely used semiconductor for thermoelectric applications, and design the inclusion to have the 300 K properties of copper, as shown in Table 1. As shown in Figure 4, the theoretical solutions well match the FEA results for a volume fraction up to 15%, with small error. With the increase of the volume fraction, the mismatch increases because the mean-field approximation becomes less valid.

We also consider the limiting case, i.e., a porous thermoelectric material (inclusion has zero electrical and thermal conductivities and Seebeck coefficient), as depicted in Figure 5. With the increase of the porosity (volume fraction of voids), the effective electrical conductivity and the effective thermal conductivity decrease at the same rate, while the Seebeck coefficient and ZT are independent of the porosity. We can easily understand the porosity-independent ZT according to the definition of the thermoelectric properties. Equation (1) is linear and holds for the zero-electric current density condition: $0 = \boldsymbol{\sigma}(\boldsymbol{e} - \boldsymbol{\alpha}\nabla T)$ or $\boldsymbol{e} = \boldsymbol{\alpha}\nabla T$. By integrating the left and right sides, we obtain the relationship between the open-circuit voltage ($V_{OC}$) and the temperature difference, as follows.

$$V_{OC} = -\int_0^L \boldsymbol{e} \cdot dx = -\int_0^L \boldsymbol{\alpha}\nabla T \cdot dx = \alpha(T_H - T_C) \tag{18}$$

Hence, $V_{OC}$ is constant when the temperature varies continuously within the material domain. Because the effective Seebeck coefficient ($\alpha_{eff}$) is the ratio of the open-circuit voltage ($V_{OC}$) to the temperature difference ($T_H - T_C$) between the hot and cold sides, $\alpha_{eff}$ is always the same regardless of the porosity. The ratio of the electrical conductivity to the thermal conductivity is maintained for a porous material. From Equation (1), with zero temperature gradient, we obtain $\frac{\boldsymbol{J_E}(\boldsymbol{\nabla}T=0)}{\boldsymbol{\sigma}} = -\boldsymbol{\nabla}\phi$. From Equation (2), with zero $J_E$, we obtain $\frac{\boldsymbol{q}(\boldsymbol{J_E}=0)}{\boldsymbol{\kappa}} = -\boldsymbol{\nabla}T$; the equations and boundary conditions for the electrical and thermal currents are exactly the same. Thus, the effective conductivities should have identical dependence on the spatial distribution of voids. The effective conductivities can be expressed as $\sigma_{eff} = B\sigma$ and $\kappa_{eff} = B\kappa$, where $B$ is the geometry factor accounting for the effect of the porosity with $B \leq 1$. Finally, because the definition of ZT is $\frac{\alpha^2\sigma}{\kappa}T$, the effective ZT of a porous material is equal to the ZT of the bulk material, and in general, the ZT is independent of the porosity.

**Effective Thermoelectric Properties in Presence of Interfacial Resistance**

The theoretical analyses presented thus far do not consider the interfacial imperfection between the matrix and the inclusion. However, a non-negligible interfacial thermal resistance exists not only for the incoherent interface between the matrix and the inclusion involving physical defects but also for the coherent interface when the phonon spectra between the two materials are very different. An interfacial electrical resistance also exists, which is affected by the coupling of the interface chemistry, contact mechanics, and charge-transport mechanism. Hence, to model realistic thermoelectric composites with a large interfacial area fraction, it is essential to consider the interfacial resistance in order to accurately predict the effective material properties of the composite. In the literature, the interfacial electrical resistance is often referred to as the electrical contact resistance, and the interfacial thermal resistance is often referred to as the Kapitza resistance. As schematically depicted in Figure 1, the interfacial electrical resistance $\beta$ and interfacial thermal resistance $\theta$ are related to the electrical potential and temperature jumps, as follows.

$$\Delta\phi = -\beta \boldsymbol{J_E} \cdot \boldsymbol{n} \tag{19}$$

$$\Delta T = -T\theta \boldsymbol{J_S} \cdot \boldsymbol{n} \tag{20}$$

When the combined potential field $\boldsymbol{U}$ for the electric potential and the temperature is defined as

$$U_H = \begin{Bmatrix} \phi & H = 1 \\ T & H = 2 \end{Bmatrix} \text{ where } \boldsymbol{Z} = -\nabla \boldsymbol{U}, \tag{21}$$

the potential jump due to interfacial resistances is expressed as

$$\Delta U_H = -\eta_{HP} J_{Ph} n_h = -\begin{bmatrix} \beta & 0 \\ 0 & T_{avg}\theta \end{bmatrix} \begin{bmatrix} J_1^e & J_2^e & J_3^e \\ J_1^s & J_2^s & J_3^s \end{bmatrix} \begin{bmatrix} n_1 \\ n_2 \\ n_3 \end{bmatrix}, \tag{22}$$

where $\boldsymbol{\eta}$ is a $2 \times 2$ matrix containing the interfacial electrical and thermal resistances in its diagonal components. In the heat conduction or electrical conduction problem, it is analytically proven that the flux within a spherical inclusion in the presence of interfacial resistance is constant.[12] Because of the mathematical analogy between the heat conduction and the thermoelectricity, the flux within a spherical inclusion in the presence of interfacial resistance is constant in the case of thermoelectricity as well. The field tensor $\boldsymbol{Z}$ within the inclusion in the presence of the interfacial resistance is given by the following equations.[12,13]

$$\begin{aligned} Z_{Rm}(x) &= \frac{\partial}{\partial x_m} \int_V \frac{\partial G_{MR}(\boldsymbol{y} - \boldsymbol{x})}{\partial y_i} E_{HiMn}\, dV(\boldsymbol{y}) Z_{Hn}^* \\ &\quad - \frac{\partial}{\partial x_m} \int_S \Delta U_H E_{HiMn} \frac{\partial G_{MR}(\boldsymbol{y} - \boldsymbol{x})}{\partial y_i} n_n\, dS(\boldsymbol{y}) \end{aligned} \tag{23}$$

$$= S_{RmHn} Z_{Hn}^* + \eta_{HP} \frac{\partial}{\partial x_m} \int_S J_{Ph} E_{HiMn} \frac{\partial G_{MR}(\boldsymbol{y} - \boldsymbol{x})}{\partial y_i} n_h(\boldsymbol{y}) n_n(\boldsymbol{y})\, dS(\boldsymbol{y}) \tag{24}$$

$$= S_{RmHn}Z^*_{Hn} + \eta_{HP}\frac{\partial}{\partial x_m}\int_S E_{PhQq}\big(Z_{Qq} - Z^*_{Qq}\big)E_{HiMn}\frac{\partial G_{MR}(\boldsymbol{y}-\boldsymbol{x})}{\partial y_i}n_h(\boldsymbol{y})n_n(\boldsymbol{y})\,dS(\boldsymbol{y}) \tag{25}$$

$$= S_{RmHn}Z^*_{Hn} + \eta_{HP}E_{PhQq}\big(Z_{Qq} - Z^*_{Qq}\big)E_{HiMn}M_{iMnhRm} \tag{26}$$

$$\text{where } M_{iMnhRm} = \frac{\partial}{\partial x_m}\int_S \frac{\partial G_{MR}(\boldsymbol{y}-\boldsymbol{x})}{\partial y_i}n_h(\boldsymbol{y})n_n(\boldsymbol{y})\,dS(\boldsymbol{y}) \tag{27}$$

Accordingly, the modified Eshelby tensor $\boldsymbol{S^M}$, which relates the eigenfield $\boldsymbol{Z}^*$ to the interior field $\boldsymbol{Z}$ in the presence of interfacial resistance, is defined as follows.

$$Z_{Qq} = S^M_{QqHn}Z^*_{Hn} \tag{28}$$

$$S^M_{QqHn} = \big(\delta_{RQ}\delta_{mq} - \eta_{HP}E_{PhQq}E_{HiMn}M_{iMnhRm}\big)^{-1}(S_{RmHn} - \eta_{KP}E_{PhHn}E_{KiMs}M_{iMshRm}) \tag{29}$$

For a spherical inclusion with radius $R$, we have

$$E_{HiMn}M_{iMnhRm} = E_{HiMn}\frac{\partial}{\partial x_m}\int_S \frac{\partial G_{MR}(\boldsymbol{y}-\boldsymbol{x})}{\partial y_i}\frac{y_h}{R}n_n(\boldsymbol{y})\,dS(\boldsymbol{y}) \tag{30}$$

$$= \frac{1}{R}\left[E_{HiMn}\frac{\partial}{\partial x_m}\int_V \frac{\partial G_{MR}(y-x)}{\partial y_n \partial y_i}y_h dV(y) + E_{HiMn}\delta_{hn}\frac{\partial}{\partial x_m}\int_V \frac{\partial G_{MR}(y-x)}{\partial y_i}dV(y)\right] \tag{31}$$

$$= \frac{1}{R}[-\delta_{HR}\delta_{mh} + S_{RmHh}], \tag{32}$$

which allows us to arrange the modified Eshelby tensor $\boldsymbol{S^M}$ as follows:

$$S^M_{QqHn} = \left[\delta_{RQ}\delta_{mq} - \frac{1}{R}\eta_{HP}E_{PhQq}(-\delta_{HR}\delta_{mh} + S_{RmHh})\right]^{-1}\left[S_{RmHn} - \frac{1}{R}\eta_{KP}E_{PhHn}(-\delta_{KR}\delta_{mh} + S_{RmKh})\right]. \tag{33}$$

Equation (33) can be simplified further for an isotropic matrix using Equation (11), as follows.

$$S_{QqHn}^{M} = \left[\delta_{RQ}\delta_{mq} + \frac{2}{3R}\eta_{RP}E_{PmQq}\right]^{-1}\left[\frac{1}{3}\delta_{RH}\delta_{mn} + \frac{2}{3R}\eta_{RP}E_{PmHn}\right] \tag{34}$$

By applying the modified Eshelby tensor within the Mori–Tanaka framework in the presence of interfacial resistance, the effective thermoelectric matrix for a composite is obtained as

$$\boldsymbol{E_{eff}} = (c_0\boldsymbol{E_0} + c_1\boldsymbol{E_1}\boldsymbol{A^M})\left(c_0\boldsymbol{I} + c_1\boldsymbol{A^M} + \frac{c_1}{R}\boldsymbol{\bar{\eta}}\boldsymbol{E_1}\boldsymbol{A^M}\right)^{-1}, \tag{35}$$

where the modified concentration tensor is $\boldsymbol{A^M} = \left(\boldsymbol{I} + \boldsymbol{S}\boldsymbol{E_0^{-1}}(\boldsymbol{E_1} - \boldsymbol{E_0}) + \frac{1}{R}\boldsymbol{\bar{\eta}}(\boldsymbol{I} - \boldsymbol{S})\boldsymbol{E_1}\right)^{-1}$, $\bar{\eta}_{RmPh} = \eta_{RP}\delta_{mh}$, and $c_0 + c_1 = 1$. The explicit $6 \times 6$ matrix representations of $\boldsymbol{E_0}$, $\boldsymbol{E_1}$, $\boldsymbol{S}$, and $\boldsymbol{\bar{\eta}}$ are presented in Supplementary Note 4. Although explicit solutions for the effective thermal and electrical conductivities and Seebeck coefficient accounting for the interfacial imperfection can be obtained from Equation (35) via the symbolic operation function implemented in a software such as Mathematica or MATLAB, we omit the closed-form expressions owing to their complexity.

The theoretical solutions are validated against FEA results based on the identical RVEs used in the previous section for a wide range of the interfacial resistances, as summarised in Figure 6, which presents the interfacial electrical and thermal resistance normalised with respect to the following values.

$$\beta_0 = \frac{R}{2\sigma_0} \tag{36}$$

$$\theta_0 = \frac{R}{2\kappa_0} \tag{37}$$

For a wide range of interfacial resistances, the theoretical predictions well match the FEA results up to an inclusion volume fraction of 15% (within the validity range of the Mori–Tanaka method). The effective thermoelectric properties approach those of a porous material in the limit of very large $\beta$ and $\theta$, while they become the values of perfect interface composites in the limit of $\beta = \theta = 0$.

The effective electrical conductivity decreases with the increase of $\beta$ at a fixed $\theta$, and it decreases with the increase of $\theta$ at a fixed $\beta$, as shown in Figure 6 (a)–(c). Clearly, the interfacial electrical resistance has a significant influence on the effective electrical conductivity. The interfacial thermal resistance also reduces the effective electrical conductivity because the electrical and heat conductions are coupled to each other in the thermoelectric materials. Similar behaviour is observed in the effective thermal conductivity, as shown in Figure 6 (d)–(f). In summary, both the electrical and heat conductivities are monotonically decreasing functions of the interfacial electrical and thermal resistances.

Interestingly, when the Seebeck coefficient of the matrix is larger than that of the inclusion, the effective Seebeck coefficient increases with the interfacial electrical resistance $\beta$ at a fixed $\theta$, as shown in Figure 6 (g) and (h). On the other hand, when the Seebeck coefficient of the matrix is smaller than that of the inclusion, the effective Seebeck coefficient decreases as $\beta$ increases for a fixed $\theta$, as follows.

$$\beta \uparrow \quad \rightarrow \quad \alpha_{eff} \uparrow \qquad (\alpha_0 > \alpha_1) \tag{38}$$

$$\beta \uparrow \quad \rightarrow \quad \alpha_{eff} \downarrow \qquad (\alpha_0 < \alpha_1) \tag{39}$$

This is because influence of the Seebeck coefficient of the inclusion becomes smaller as $\beta$ increases; thus, the effective Seebeck coefficient converges to that of the porous matrix (which is identical to that of the pure bulk matrix, as discussed previously). That is, a sufficiently large electrical contact resistance ($\beta/\beta_O \gg 1$) blocks the electrical current through the inclusion, causing the inclusion to behave as an electrically insulating medium. On the other hand, the effective Seebeck coefficient decreases as $\theta$ increases regardless of the relative sizes of the Seebeck coefficients of the matrix and inclusions. Figure 6 (g) and (i) show the effective Seebeck coefficient when the Seebeck coefficient of the matrix is larger than that of the inclusion.

$$\theta \uparrow \quad \rightarrow \quad \alpha_{eff} \downarrow \qquad (\alpha_0 > \alpha_1) \tag{40}$$

$$\theta \uparrow \quad \rightarrow \quad \alpha_{eff} \downarrow \qquad (\alpha_0 < \alpha_1) \tag{41}$$

This can be understood from the definition of the Seebeck coefficient. The Seebeck coefficient refers to the generated voltage difference given the temperature difference through the electrical path. If a temperature jump occurs at the interface, the effective total temperature difference and the $V_{OC}$ are reduced across the hot to cold side, while the electrical current can flow through the inclusion. As a result, the thermoelectric composite produces less voltage as $\theta$ increases for a given temperature difference across the hot and cold sides.

Having obtained the effective electrical and thermal conductivities and Seebeck coefficient analytically and numerically, we can obtain the ZT using Equation (17) in terms of the interfacial electrical and thermal resistances, as shown in Figure 7. The upper right corner of the two-dimensional ZT map corresponds to the porous material limit, and the lower left corner corresponds to the perfect interface limit. As discussed in the previous section, the ZT of the porous materials is identical to that of the pure matrix because the effects of the porosity on the electrical and thermal conductivities cancel each other out, while the Seebeck coefficient is independent of the porosity. Additionally, the ZT of the composite decreases when either the interfacial electrical resistance or the interfacial thermal resistance is significantly larger than the other. The theoretically predicted ZTs of thermoelectric composites having a few other combinations of the matrix and inclusion materials are presented in Supplementary Figure 1.

**Evaluation of Effective Thermoelectric Properties of Composite Subjected to Large Temperature Difference**

Thus far, for the ease of theoretical derivation, we assumed that the temperature difference between the hot and cold sides of the thermoelectric composite is very small. However, in practice, thermoelectric composites are subjected to a large temperature gradient on the order of $100 \ \mathrm{K/mm}$. Therefore, it is necessary to consider the effective properties of a thermoelectric composite subjected to a large temperature difference. As stated in the Introduction, we do not consider the temperature-dependent thermoelectric property in this study; rather, we focus on the explicit *T* dependence in the governing equation, as well as the thermoelectric matrix.

The closed-form solutions for the thermoelectric properties obtained from Equation (14) include many terms involving the average temperature $T_{avg}$ (see Supplementary Note 3 for the explicit equation). All terms with $T_{avg}$ have multiplicative factors containing square- or cubic-order Seebeck coefficients of the matrix and/or the inclusion. For realistic thermoelectric materials, these terms are negligible, as the Seebeck coefficient is on the order of $10^{-6}$ in SI units, while the operating temperature of a thermoelectric device is less than or equal to $10^{3}$ K. Hence, the effective thermoelectric properties hardly depend on the temperature when the temperature dependence of the material properties of the matrix and the inclusion are not considered. This can be confirmed by comparing the theoretical results obtained using different $T_{avg}$ values as the input, as shown in Figure 8 (a) and (b).

We now consider a thermoelectric composite subjected to a large temperature difference $(T_{\mathrm{H}} - T_{C})$ that can be up to several hundreds of Kelvin. This thermoelectric composite can be cut into many pieces so that the thermoelectric composite is regarded as a serial connection of pieces with a small temperature difference, as schematically illustrated in Figure 8 (c). As mentioned previously, the pieces of the composite have almost identical thermoelectric properties, although their average temperatures are very different. Hence, one can expect that a thermoelectric composite subjected to a large temperature difference would behave similarly to a thermoelectric composite subjected to a small temperature difference. To confirm this conjecture, we compare our theoretical prediction with FEA results obtained with a relatively large temperature difference, i.e., a relatively large $\frac{T_{\mathrm{H}}-T_C}{T_{avg}}$. A good match is obtained, as depicted in Figure 8 (d).

**Conclusion**

The theoretical solution for the effective thermoelectric properties of a composite is obtained for a composite involving spherical inclusions with interfacial electrical and thermal resistances. Our prediction is validated against FEA calculations as well as the limiting case with known behaviour: a porous thermoelectric material. In addition, we show that the theoretical solution can be applied to predict the properties of a composite subjected to a large temperature difference, despite the explicit $T$ dependence in the governing equation.

Although we made an important step towards the prediction of the effective thermoelectric property, there remain a few critical challenges. Realistic thermoelectric materials have non-negligible anisotropy, and the thermoelectric properties have significant temperature dependence within the operating temperature range. The anisotropy can be relatively easily (but not trivially) treated by the numerical integration of the Eshelby tensor. However, the non-monotonic temperature-dependent thermoelectric properties must be handled via a careful nonlinear homogenisation framework. Overcoming these challenges will be the objective of our continuing research endeavours. We believe that this line of study will pave the way for the systematic development of thermoelectric composites with enhanced efficiency.

**Tables**

| | $Bi_2Te_3$ | Copper |
|---|---|---|
| Thermal conductivity $(\kappa)$ [W/mK] | 2 | 400 |
| Electrical conductivity $(\sigma)$ [S/m] | $5 \times 10^4$ | $6 \times 10^7$ |
| Seebeck coefficient $(\alpha)$ [μV/K] | 230 | 6.5 |
| ZT for 300 K | 0.397 | 0.0019 |

Table 1. Material properties used for comparing the theoretical solution with the FEA results. The material properties of $Bi_2Te_3$ at 300 K are used for the matrix.[14] The material properties of copper are used for the inclusions.[15]

**Figures and captions**

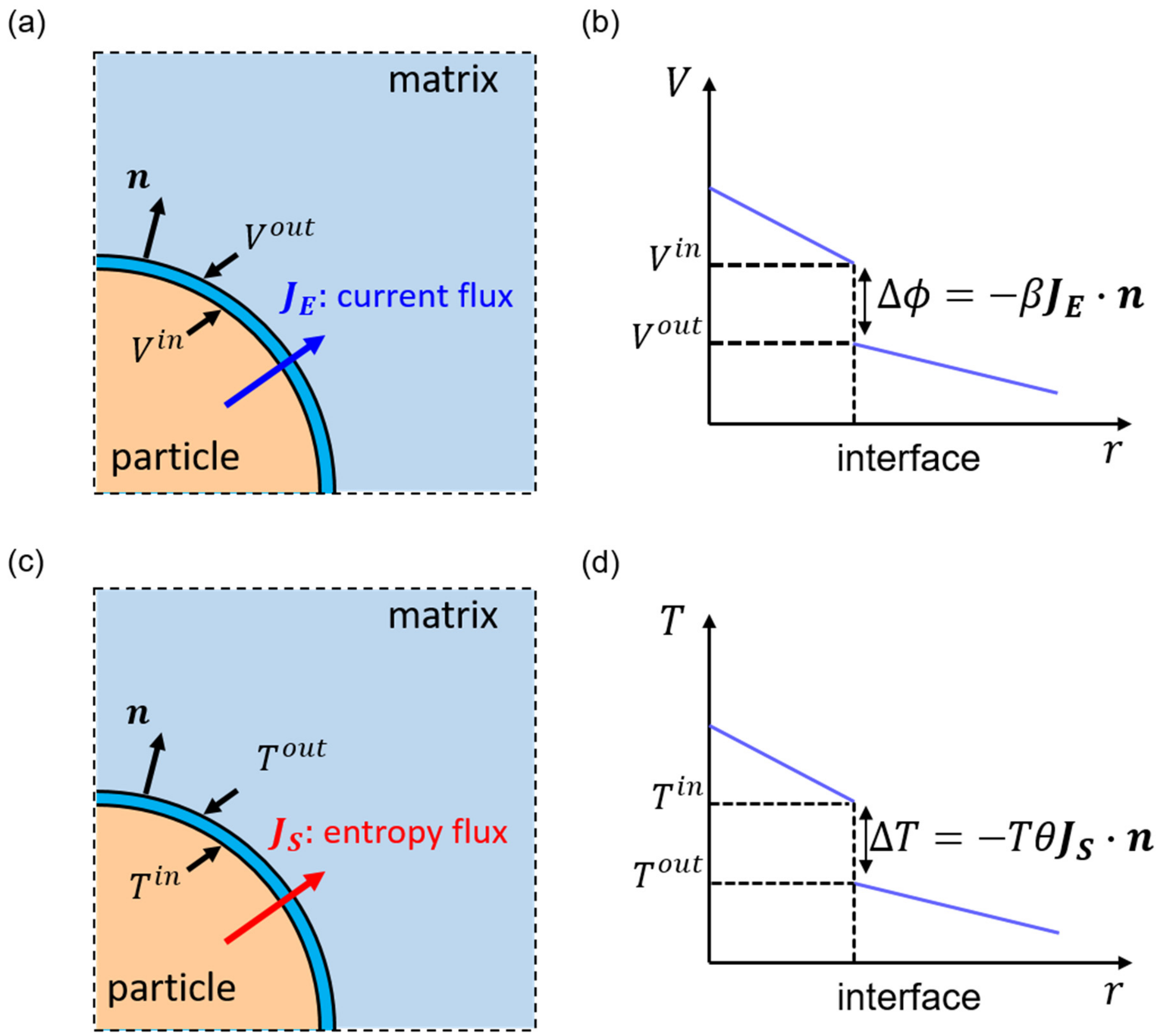

Figure 1. (a) Configuration of the interfacial electrical resistance. (b) Voltage jump through the interfacial electrical resistance. (c) Configuration of the interfacial thermal resistance. (d) Temperature jump through the interfacial thermal resistance.

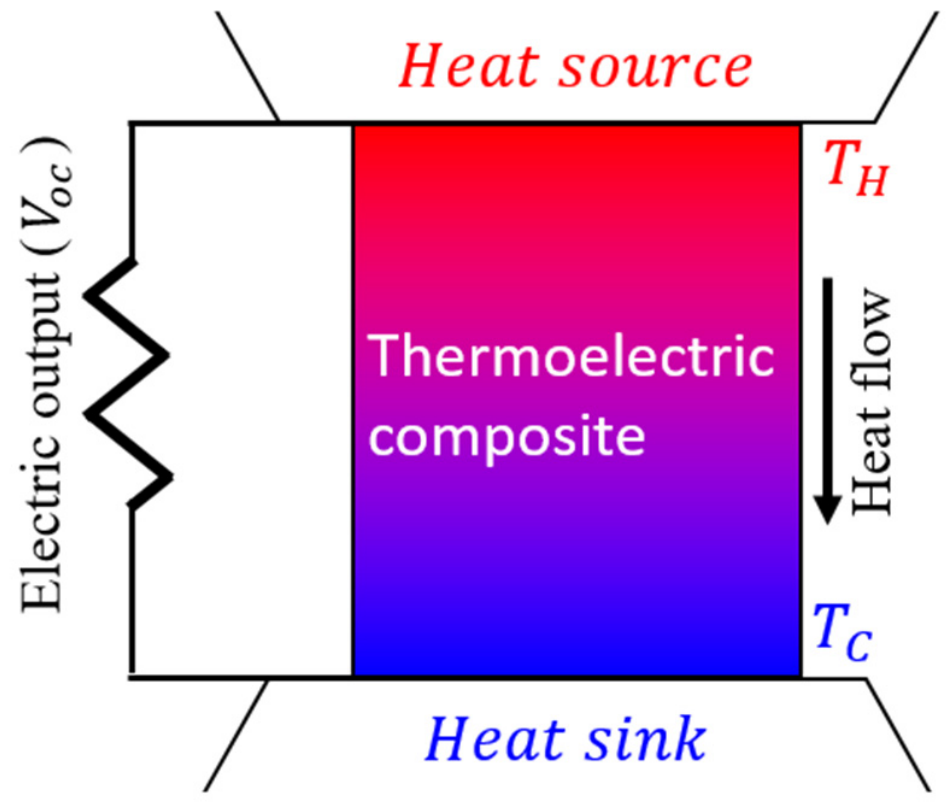

Figure 2. Configuration of the boundary condition for the thermoelectric generator.

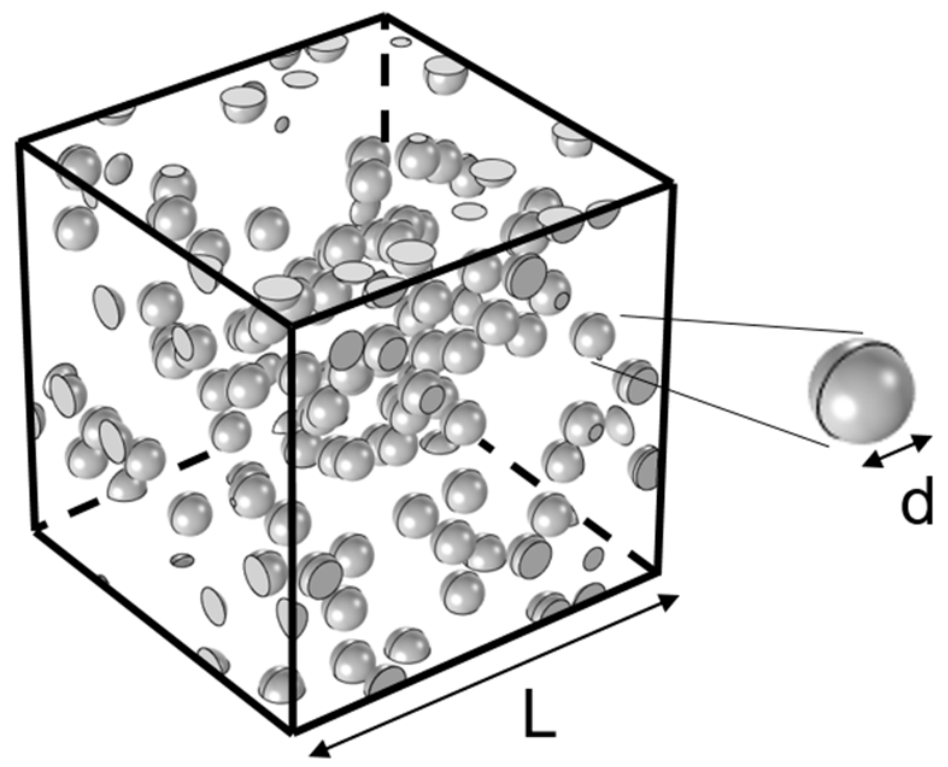

Figure 3. Configuration of RVE for the FEA at a volume fraction of 10%. The matrix size is 10 mm × 10 mm × 10 mm, and the diameter of the inhomogeneity is 1 mm (i.e., d = 1 mm and L = 10 mm).

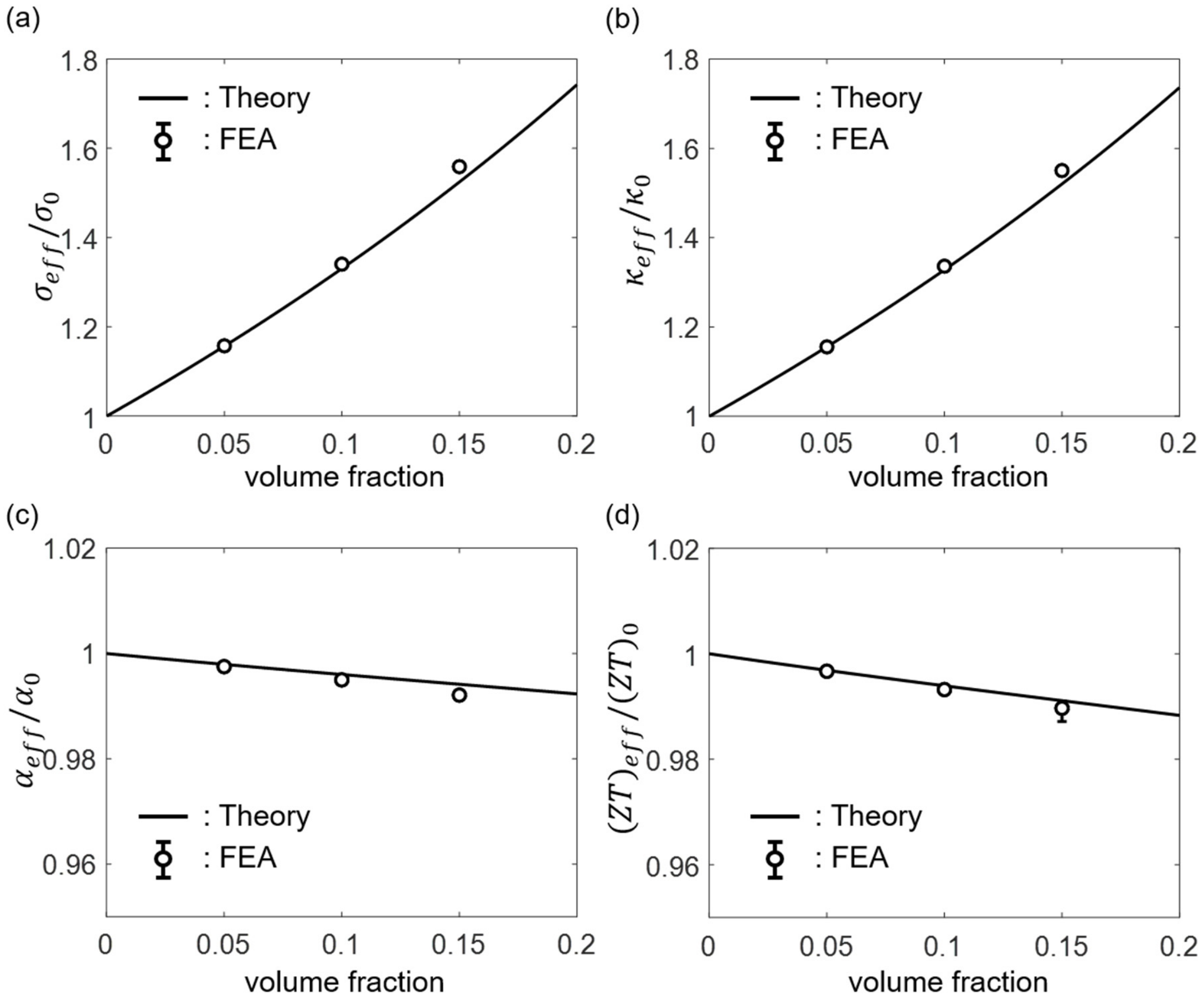

Figure 4. Normalised effective thermoelectric properties and ZT ((a)$\sigma_{eff}/\sigma_0$, (b)$\kappa_{eff}/\kappa_0$, (c)$\alpha_{eff}/\alpha_0$, (d)$(ZT)_{eff}/(ZT)_0$) for a composite consisting of a $Bi_2Te_3$ matrix and copper inclusion, where the subscript 0 refers to the properties of the matrix. The error bars are covered by markers.

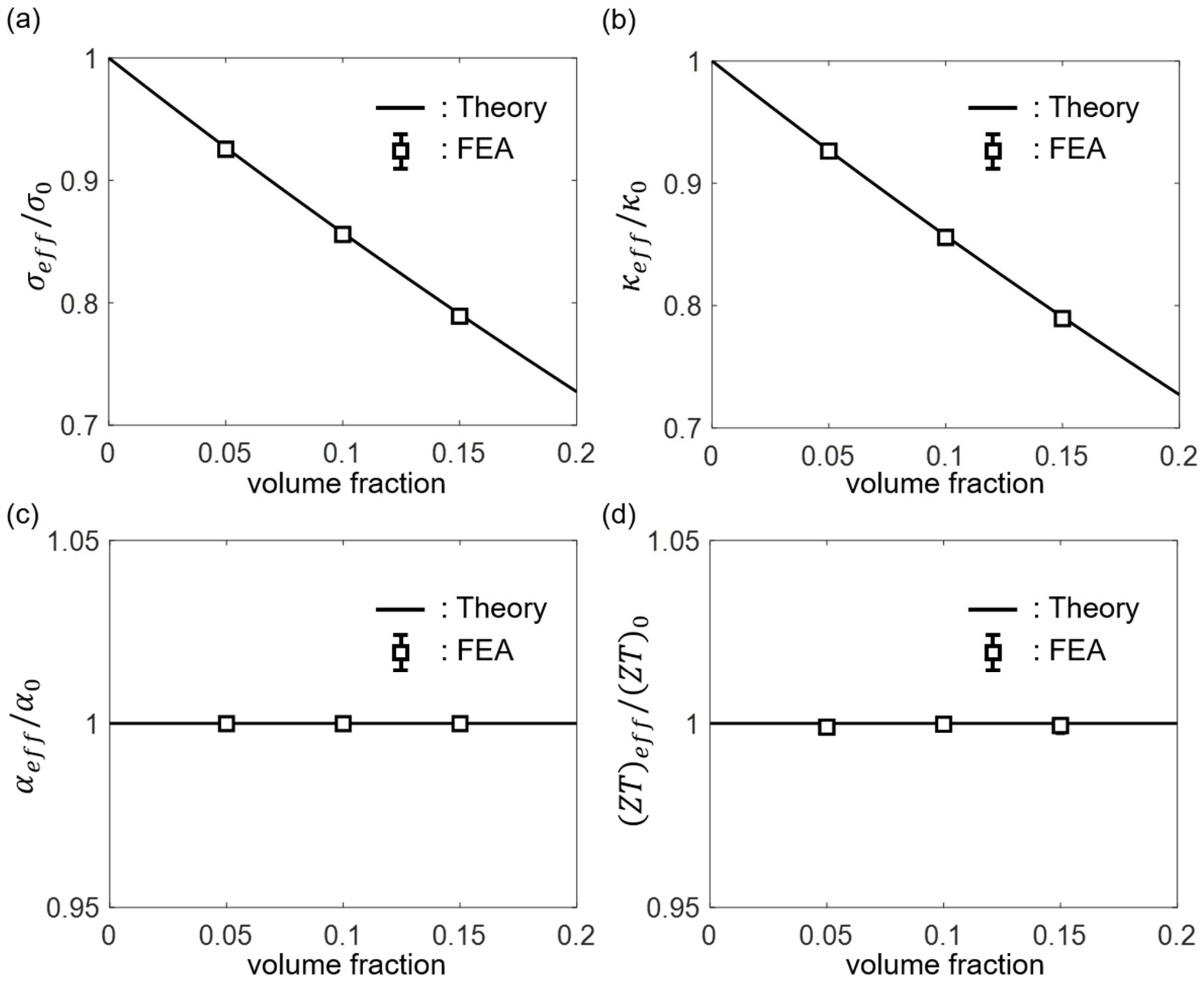

Figure 5. Normalised effective thermoelectric properties and effective ZT for a porous medium ((a) $\sigma_{eff}/\sigma_0$, (b) $\kappa_{eff}/\kappa_0$, (c) $\alpha_{eff}/\alpha_0$, (d) $(ZT)_{eff}/(ZT)_0$), where the subscript 0 refers to the properties of the bulk material. The error bars are covered by markers.

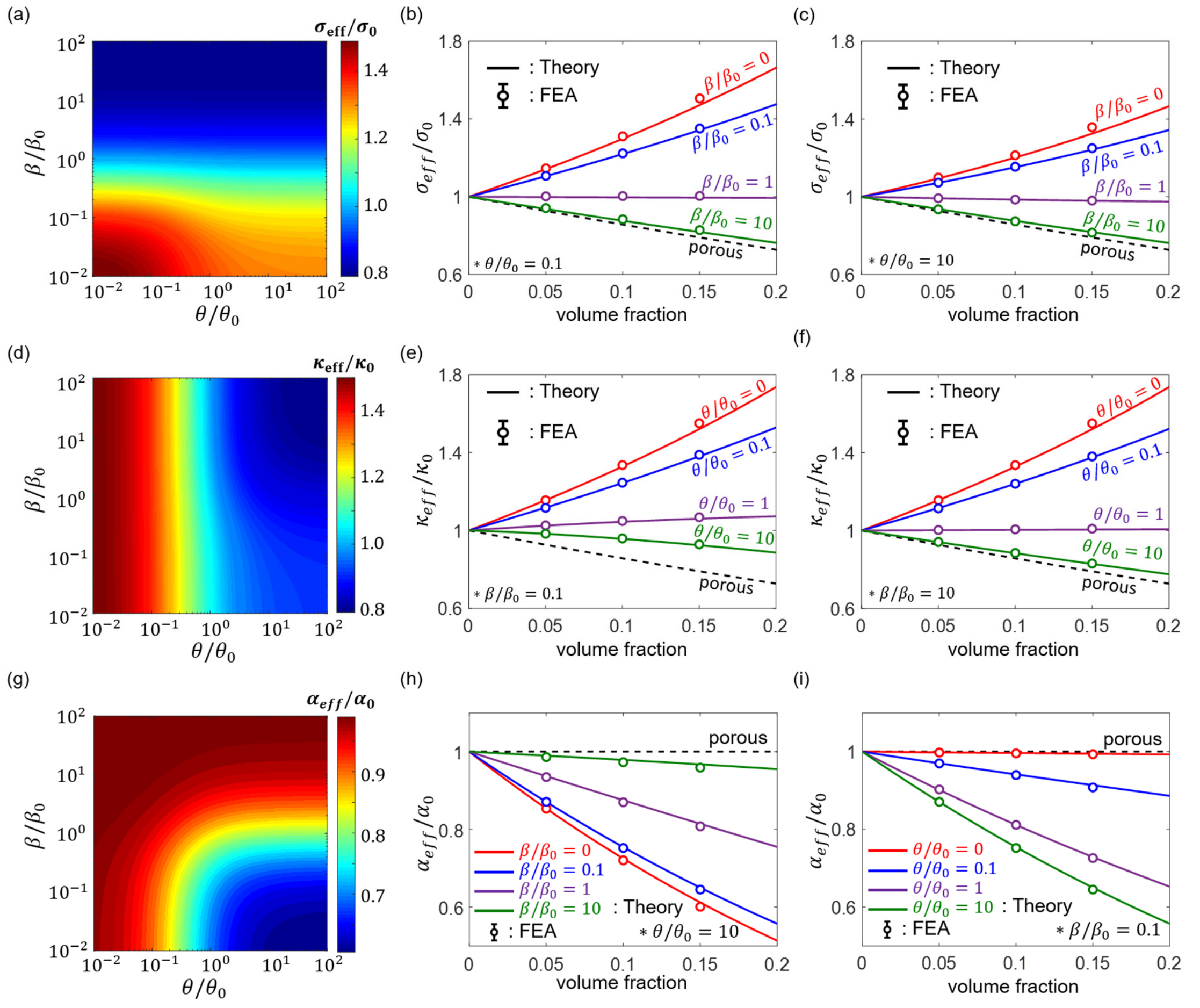

Figure 6. Normalised effective thermoelectric properties for a composite consisting of a $Bi_2Te_3$ matrix and copper inclusion in the presence of interfacial resistances. (a) $\sigma_{eff}/\sigma_0$ contour at a volume fraction of 15% in terms of $\beta$ and $\theta$. (b, c) $\sigma_{eff}/\sigma_0$ when $\theta/\theta_0 =$ 0.1 and 10, respectively. (d) $\kappa_{eff}/\kappa_0$ contour at a volume fraction of 15% in terms of $\beta$ and $\theta$. (e, f) $\kappa_{eff}/\kappa_0$ when $\beta/\beta_0 = 0.1$ and 10, respectively. (g) $\alpha_{eff}/\alpha_0$ contour at a volume fraction of 15% in terms of $\beta$ and $\theta$. (h) $\alpha_{eff}/\alpha_0$ when $\theta/\theta_0 = 10$. (i) $\alpha_{eff}/\alpha_0$ when $\beta/\beta_0 = 0.1$. The error bars are covered by markers.

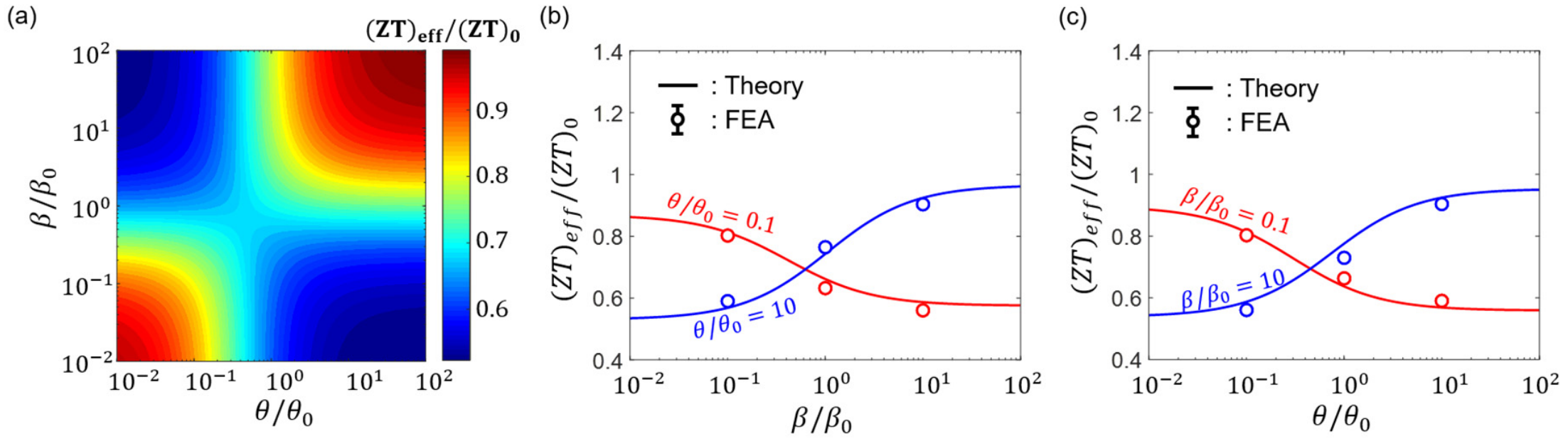

Figure 7. (a) Normalised effective ZT for a composite with a $Bi_2Te_3$ matrix and copper inclusion with respect to $\beta$ and $\theta$ at a volume fraction of 15%. (b) Normalised effective ZT with respect to $\beta/\beta_0$ compared with the FEA results. (c) Normalised effective ZT with respect to $\theta/\theta_0$ compared with the FEA results.

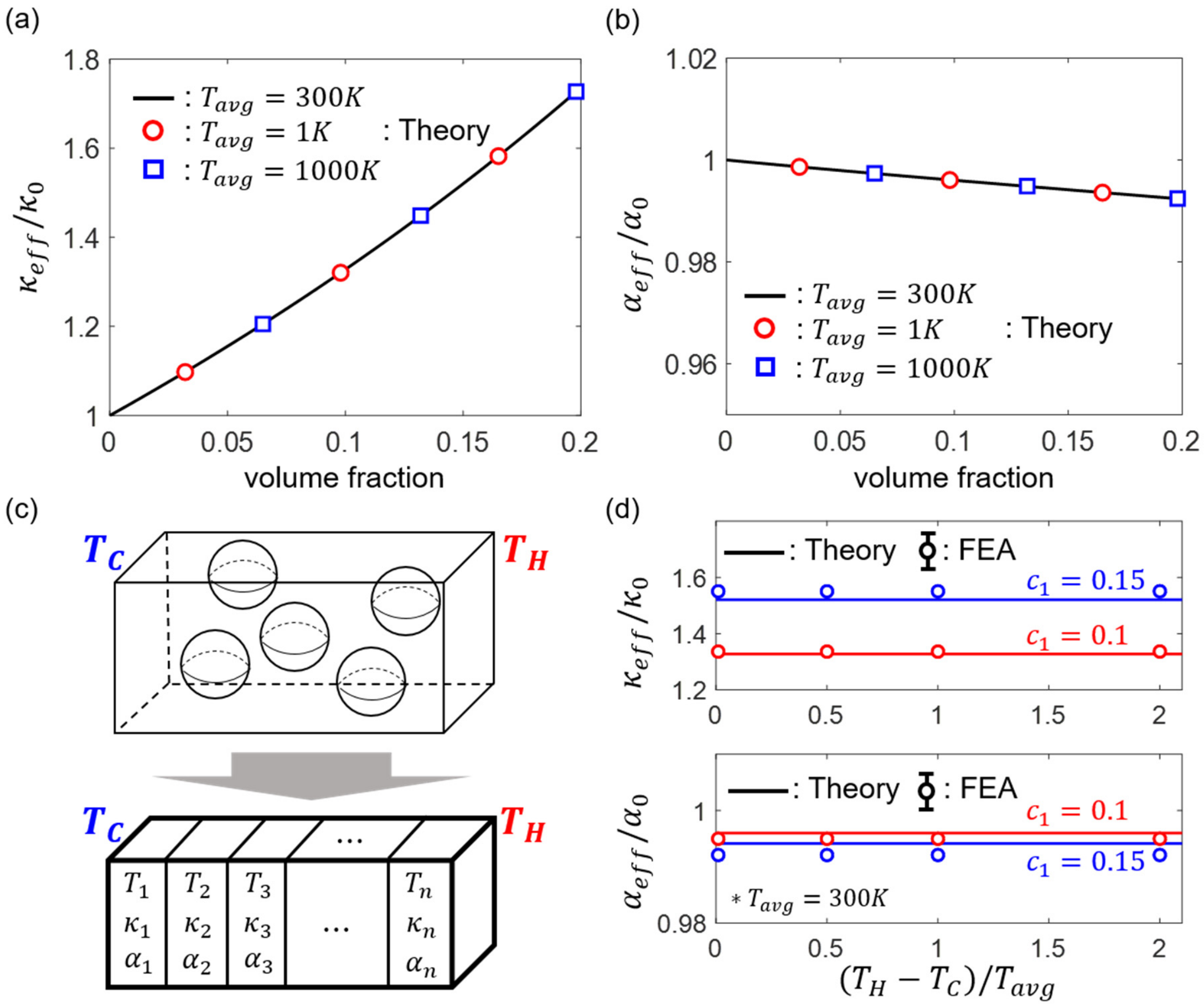

Figure 8. (a) Theoretically predicted $\kappa_{eff}/\kappa_0$ with different average temperatures as the input for a composite with a $Bi_2Te_3$ matrix and copper inclusion. (b) Theoretical $\alpha_{eff}/\alpha_0$ at different mean temperatures. (c) Configuration for converting the composite into a serial connection of homogenised pieces when a large temperature difference is imposed. (d) $\kappa_{eff}/\kappa_0$ and $\alpha_{eff}/\alpha_0$ when a large temperature difference is imposed. The error bars are covered by markers.

**Acknowledgements**

This research was supported by Basic Science Research Program (2016R1C1B2011979) and Creative Materials Discovery Program (2016M3D1A1900038) through the National Research Foundation of Korea (NRF) funded by the Ministry of Science and ICT. We also acknowledge the support from the Korea Institute of Energy Technology Evaluation and Planning (KETEP) and the Ministry of Trade Industry & Energy (MOTIE) of the Republic of Korea (No. 20162000000910). It is also supported by by Korea Electrotechnology Research Institute (KERI) Primary research program (No. 18-12-N0101-34) through the National Research Council of Science & Technology (NST) funded by the Ministry of Science and ICT.

**Author contributions Statement**

J.J. and S.R. designed the research, interpret the results, and wrote the manuscript. J.J. and S.L. carried out the analytic derivation, and J.J. carried out numerical simulations. B.R. discussed and analysed the results. All authors revised the manuscript.

**Supplementary information**

# Investigation of effective thermoelectric properties of composite with interfacial resistance using micromechanics-based homogenisation

Jiyoung Jung[1], Sangryun Lee[1], Byungki Ryu[2], and Seunghwa Ryu[1,*]

**Affiliations**

[1]Department of Mechanical Engineering, Korea Advanced Institute of Science and Technology (KAIST), 291 Daehak-ro, Yuseong-gu, Daejeon 34141, Republic of Korea

[2]Energy Conversion Research Center, Korea Electrotechnology Research Institute (KERI), 12 Bulmosan-ro 10beon-gil, Seongsan-gu, Changwon-si, Gyoengsangnam-do 51543, Republic of Korea

[*]Corresponding author e-mail: ryush@kaist.ac.kr

**Supplementary Note 1: Eshelby tensor for spherical inclusion with radius *R* in isotropic matrix**

The Eshelby tensor in thermoelectricity for an ellipsoidal inclusion is as follows.

$$S_{MnAb} = \frac{1}{4\pi} E_{PiAb} \int_{-1}^{1} \int_{0}^{2\pi} \left(\frac{\xi}{a}\right)_i \left(\frac{\xi}{a}\right)_n K_{MP}^{-1} d\theta d\xi_3$$

$$\text{where } K_{MP} = \left(\frac{\xi}{a}\right)_s \left(\frac{\xi}{a}\right)_t E_{MsPt}, \qquad \frac{x_1^2}{a_1^2} + \frac{x_2^2}{a_2^2} + \frac{x_3^2}{a_3^2} = 1$$

$$\xi_1 = (1 - \xi_3^2)^{1/2} \cos\theta\,,\ \xi_2 = (1 - \xi_3^2)^{1/2} \sin\theta\,,\ \xi_3 = \xi_3$$

For a spherical inclusion with radius *R* in the isotropic matrix, the Eshelby tensor is simplified as follows.

$$K_{MP} = \frac{\xi_s \xi_t}{R^2} E_{MsPt} \qquad \text{(for spherical inclusion)}$$

$$= \frac{1}{R^2} D_{MP} \qquad where\ \boldsymbol{D} = \begin{bmatrix} \sigma & \sigma\alpha \\ \sigma\alpha & \gamma/T \end{bmatrix}\ \&\ \xi_1^2 + \xi_2^2 + \xi_3^2 = 1 \qquad \text{(for isotropic matrix)}$$

$$K_{MP}^{-1} = R^2 (D^{-1})_{MP} \qquad where\ \boldsymbol{D^{-1}} = \frac{1}{\sigma\gamma/T - (\sigma\alpha)^2} \begin{bmatrix} \gamma/T & -\sigma\alpha \\ -\sigma\alpha & \sigma \end{bmatrix}$$

Then, $K_{MP}^{-1}$ is independent of $\xi_3$, and the Eshelby tensor is arranged as

$$S_{MnAb} = \frac{1}{4\pi R^2} (K^{-1})_{MP} E_{PiAb} \int_{-1}^{1} \int_{0}^{2\pi} \xi_i \xi_n d\theta d\xi_3$$

$$\text{because } E_{PiAb} = 0 \text{ if } \mathrm{i} \neq \mathrm{b} \text{ (isotropic), and } \int_{-1}^{1} \int_{0}^{2\pi} \xi_i \xi_n d\theta d\xi_3 = \begin{cases} \frac{4\pi}{3} & for\ \ i = n \\ 0 & for\ \ i \neq \mathrm{n} \end{cases}$$

$$S_{MnAb} = \frac{1}{3R^2} (K^{-1})_{MP} D_{PA} \delta_{\mathrm{in}}$$

$$= \frac{1}{3} \delta_{\mathrm{MA}} \delta_{\mathrm{in}}.$$

**Supplementary Note 2: Derivation of effective thermoelectric matrix $\boldsymbol{E_{eff}}$**

From $\bar{\boldsymbol{J}} = \frac{1}{\Omega}\int_{\Omega} \boldsymbol{J} d\Omega$ *and* $\bar{\boldsymbol{Z}} = \frac{1}{\Omega}\int_{\Omega} \boldsymbol{Z} d\Omega$ where $\Omega$ is the volume of the composite, we have $\bar{\boldsymbol{Z}} = c_0\overline{\boldsymbol{Z_0}} + c_1\overline{\boldsymbol{Z_1}}$ and $\bar{\boldsymbol{J}} = c_0\overline{\boldsymbol{J_0}} + c_1\overline{\boldsymbol{J_1}} = c_0\boldsymbol{E_0}\overline{\boldsymbol{Z_0}} + c_1\boldsymbol{E_1}\overline{\boldsymbol{Z_1}} = \boldsymbol{E_0}\bar{\boldsymbol{Z}} + c_1(\boldsymbol{E_1} - \boldsymbol{E_0})\overline{\boldsymbol{Z_1}}$.

From Equation (13),

$$\bar{\boldsymbol{J}} = [\boldsymbol{E_0} + c_1(\boldsymbol{E_1} - \boldsymbol{E_0})\boldsymbol{A}(c_0\boldsymbol{I} + c_1\boldsymbol{A})^{-1}]\bar{\boldsymbol{Z}},$$

where $\bar{\boldsymbol{J}} = \boldsymbol{E_{eff}}\bar{\boldsymbol{Z}}$.

$$\begin{aligned}
\boldsymbol{E_{eff}} &= \boldsymbol{E_0} + c_1(\boldsymbol{E_1} - \boldsymbol{E_0})\boldsymbol{A}(c_0\boldsymbol{I} + c_1\boldsymbol{A})^{-1} \\
&= \boldsymbol{E_0} + c_1\boldsymbol{E_1}\boldsymbol{A}(c_0\boldsymbol{I} + c_1\boldsymbol{A})^{-1} - c_1\boldsymbol{E_0}\boldsymbol{A}(c_0\boldsymbol{I} + c_1\boldsymbol{A})^{-1} \\
&= \boldsymbol{E_0}(c_0\boldsymbol{I} + c_1\boldsymbol{A})(c_0\boldsymbol{I} + c_1\boldsymbol{A})^{-1} + c_1\boldsymbol{E_1}\boldsymbol{A}(c_0\boldsymbol{I} + c_1\boldsymbol{A})^{-1} - c_1\boldsymbol{E_0}\boldsymbol{A}(c_0\boldsymbol{I} + c_1\boldsymbol{A})^{-1} \\
&= c_0\boldsymbol{E_0}(c_0\boldsymbol{I} + c_1\boldsymbol{A})^{-1} + c_1\boldsymbol{E_1}\boldsymbol{A}(c_0\boldsymbol{I} + c_1\boldsymbol{A})^{-1} \\
&= (c_0\boldsymbol{E_0} + c_1\boldsymbol{E_1}\boldsymbol{A})(c_0\boldsymbol{I} + c_1\boldsymbol{A})^{\boldsymbol{-1}}
\end{aligned}$$

or

$$\begin{aligned}
\boldsymbol{E_{eff}} &= \boldsymbol{E_0} + c_1(\boldsymbol{E_1} - \boldsymbol{E_0})\boldsymbol{A}(c_0\boldsymbol{I} + c_1\boldsymbol{A})^{-1} \\
&= \boldsymbol{E_0} + c_1[(c_0\boldsymbol{I} + c_1\boldsymbol{A})\boldsymbol{A}^{-1}(\boldsymbol{E_1} - \boldsymbol{E_0})^{-1}]^{-1} \\
&= \boldsymbol{E_0} + c_1[(c_0\boldsymbol{A}^{-1} + c_1\boldsymbol{I})(\boldsymbol{E_1} - \boldsymbol{E_0})^{-1}]^{-1} \\
&= \boldsymbol{E_0} + c_1[(\boldsymbol{I} + c_0\boldsymbol{S}\boldsymbol{E_0^{-1}}(\boldsymbol{E_1} - \boldsymbol{E_0}))(\boldsymbol{E_1} - \boldsymbol{E_0})^{-1}]^{-1} \\
&= \boldsymbol{E_0} + c_1[(\boldsymbol{E_1} - \boldsymbol{E_0})^{\boldsymbol{-1}} + c_0\boldsymbol{S}\boldsymbol{E_0^{-1}}]^{-1}
\end{aligned}$$

**Supplementary Note 3: Closed-form solutions for effective thermoelectric properties**

The closed-form solutions for the effective Seebeck coefficient and the effective electrical and thermal conductivities for a composite consisting of isotropic materials and spherical inclusions are as follows.

$$\sigma_{eff} = \sigma_0 - (3c_1\sigma_0(2\kappa_0\sigma_0 - 2\kappa_0\sigma_1 + \kappa_1\sigma_0 - \kappa_1\sigma_1 + c_1\kappa_0\sigma_0 - c_1\kappa_0\sigma_1 - c_1\kappa_1\sigma_0 + c_1\kappa_1\sigma_1 + T_{avg}\alpha_0^2\sigma_0\sigma_1 +$$
$$T_{avg}\alpha_1^2\sigma_0\sigma_1 - 2T_{AVG}\alpha_0\alpha_1\sigma_0\sigma_1 - T_{avg}\alpha_0^2c_1\sigma_0\sigma_1 - T_{avg}\alpha_1^2c_1\sigma_0\sigma_1 + 2T_{avg}\alpha_0\alpha_1c_1\sigma_0\sigma_1))/(4\kappa_0\sigma_0 + 2\kappa_0\sigma_1 +$$
$$2\kappa_1\sigma_0 + \kappa_1\sigma_1 + c_1^2\kappa_0\sigma_0 - c_1^2\kappa_0\sigma_1 - c_1^2\kappa_1\sigma_0 + c_1^2\kappa_1\sigma_1 + 4c_1\kappa_0\sigma_0 - c_1\kappa_0\sigma_1 - c_1\kappa_1\sigma_0 - 2c_1\kappa_1\sigma_1 +$$
$$2T_{avg}\alpha_0^2\sigma_0\sigma_1 + 2T_{avg}\alpha_1^2\sigma_0\sigma_1 - T_{avg}\alpha_0^2c_1^2\sigma_0\sigma_1 - T_{avg}\alpha_1^2c_1^2\sigma_0\sigma_1 - 4T_{avg}\alpha_0\alpha_1\sigma_0\sigma_1 - T_{avg}\alpha_0^2c_1\sigma_0\sigma_1 -$$
$$T_{avg}\alpha_1^2c_1\sigma_0\sigma_1 + 2T_{avg}\alpha_0\alpha_1c_1^2\sigma_0\sigma_1 + 2T_{avg}\alpha_0\alpha_1c_1\sigma_0\sigma_1)$$

$$\kappa_{eff} = (\kappa_0(4\kappa_0\sigma_0 + 2\kappa_0\sigma_1 + 2\kappa_1\sigma_0 + \kappa_1\sigma_1 + 4c_1^2\kappa_0\sigma_0 - 4c_1^2\kappa_0\sigma_1 - 4c_1^2\kappa_1\sigma_0 + 4c_1^2\kappa_1\sigma_1 - 8c_1\kappa_0\sigma_0$$
$$+2c_1\kappa_0\sigma_1 + 2c_1\kappa_1\sigma_0 + 4c_1\kappa_1\sigma_1 + 2T_{avg}\alpha_0^2\sigma_0\sigma_1 + 2T_{avg}\alpha_1^2\sigma_0\sigma_1 - 4T_{avg}\alpha_0^2c_1^2\sigma_0\sigma_1 - 4T_{avg}\alpha_1^2c_1^2\sigma_0\sigma_1$$
$$-4T_{avg}\alpha_0\alpha_1\sigma_0\sigma_1 + 2T_{avg}\alpha_0^2c_1\sigma_0\sigma_1 + 2T_{avg}\alpha_1^2c_1\sigma_0\sigma_1 + 8T_{avg}\alpha_0\alpha_1c_1^2\sigma_0\sigma_1 - 4T_{avg}\alpha_0\alpha_1c_1\sigma_0\sigma_1))/(4\kappa_0\sigma_0$$
$$+2\kappa_0\sigma_1 + 2\kappa_1\sigma_0 + \kappa_1\sigma_1 - 2c_1^2\kappa_0\sigma_0 + 2c_1^2\kappa_0\sigma_1 + 2c_1^2\kappa_1\sigma_0 - 2c_1^2\kappa_1\sigma_1 - 2c_1\kappa_0\sigma_0 + 5c_1\kappa_0\sigma_1 - 4c_1\kappa_1\sigma_0$$
$$+c_1\kappa_1\sigma_1 + 2T_{avg}\alpha_0^2\sigma_0\sigma_1 + 2T_{avg}\alpha_1^2\sigma_0\sigma_1 + 2T_{avg}\alpha_0^2c_1^2\sigma_0\sigma_1 + 2T_{avg}\alpha_1^2c_1^2\sigma_0\sigma_1 - 4T_{avg}\alpha_0\alpha_1\sigma_0\sigma_1$$
$$-4T_{avg}\alpha_0^2c_1\sigma_0\sigma_1 - 4T_{avg}\alpha_1^2c_1\sigma_0\sigma_1 - 4T_{avg}\alpha_0\alpha_1c_1^2\sigma_0\sigma_1 + 8T_{avg}\alpha_0\alpha_1c_1\sigma_0\sigma_1)$$

$$\alpha_{eff} = (4\alpha_0\kappa_0\sigma_0 + 2\alpha_0\kappa_0\sigma_1 + 2\alpha_0\kappa_1\sigma_0 + \alpha_0\kappa_1\sigma_1 - 2\alpha_0c_1\kappa_0\sigma_0 - 4\alpha_0c_1\kappa_0\sigma_1 - 4\alpha_0c_1\kappa_1\sigma_0 + \alpha_0c_1\kappa_1\sigma_1$$
$$+9\alpha_1c_1\kappa_0\sigma_1 + 2T_{avg}\alpha_0^3\sigma_0\sigma_1 - 2\alpha_0c_1^2\kappa_0\sigma_0 + 2\alpha_0c_1^2\kappa_0\sigma_1 + 2\alpha_0c_1^2\kappa_1\sigma_0 - 2\alpha_0c_1^2\kappa_1\sigma_1 + 2T_{avg}\alpha_0^3c_1^2\sigma_0\sigma_1$$
$$+2T_{avg}\alpha_0\alpha_1^2\sigma_0\sigma_1 - 4T_{avg}\alpha_0^2\alpha_1\sigma_0\sigma_1 - 4T_{avg}\alpha_0^3c_1\sigma_0\sigma_1 - 4T_{avg}\alpha_0\alpha_1^2c_1\sigma_0\sigma_1 + 8T_{avg}\alpha_0^2\alpha_1c_1\sigma_0\sigma_1$$
$$+2T_{avg}\alpha_0\alpha_1^2c_1^2\sigma_0\sigma_1 - 4T_{avg}\alpha_0^2\alpha_1c_1^2\sigma_0\sigma_1)/(4\kappa_0\sigma_0 + 2\kappa_0\sigma_1 + 2\kappa_1\sigma_0 + \kappa_1\sigma_1 - 2c_1^2\kappa_0\sigma_0 + 2c_1^2\kappa_0\sigma_1 + 2c_1^2\kappa_1\sigma_0$$
$$-2c_1^2\kappa_1\sigma_1 - 2c_1\kappa_0\sigma_0 + 5c_1\kappa_0\sigma_1 - 4c_1\kappa_1\sigma_0 + c_1\kappa_1\sigma_1 + 2T_{avg}\alpha_0^2\sigma_0\sigma_1 + 2T_{avg}\alpha_1^2\sigma_0\sigma_1 + 2T_{avg}\alpha_0^2c_1^2\sigma_0\sigma_1$$
$$+2T_{avg}\alpha_1^2c_1^2\sigma_0\sigma_1 - 4T_{avg}\alpha_0\alpha_1\sigma_0\sigma_1 - 4T_{avg}\alpha_0^2c_1\sigma_0\sigma_1 - 4T_{avg}\alpha_1^2c_1\sigma_0\sigma_1 - 4T_{avg}\alpha_0\alpha_1c_1^2\sigma_0\sigma_1$$
$$+8T_{avg}\alpha_0\alpha_1c_1\sigma_0\sigma_1)$$

**Supplementary Note 4: Explicit expressions for $\boldsymbol{E_0}, \boldsymbol{E_1}, \boldsymbol{S}, and\ \boldsymbol{\bar{\eta}}$**

With thermoelectric properties for the matrix $\boldsymbol{\sigma}^{(0)}, \boldsymbol{\kappa}^{(0)}, \boldsymbol{\alpha}^{(0)}$

$$\boldsymbol{E_0} = \begin{bmatrix} \sigma_{11}^{(0)} & \sigma_{12}^{(0)} & \sigma_{13}^{(0)} & \sigma_{1k}^{(0)}\alpha_{k1}^{(0)} & \sigma_{1k}^{(0)}\alpha_{k2}^{(0)} & \sigma_{1k}^{(0)}\alpha_{k3}^{(0)} \\ \sigma_{21}^{(0)} & \sigma_{22}^{(0)} & \sigma_{23}^{(0)} & \sigma_{2k}^{(0)}\alpha_{k1}^{(0)} & \sigma_{2k}^{(0)}\alpha_{k2}^{(0)} & \sigma_{2k}^{(0)}\alpha_{k3}^{(0)} \\ \sigma_{31}^{(0)} & \sigma_{32}^{(0)} & \sigma_{33}^{(0)} & \sigma_{3k}^{(0)}\alpha_{k1}^{(0)} & \sigma_{3k}^{(0)}\alpha_{k2}^{(0)} & \sigma_{3k}^{(0)}\alpha_{k3}^{(0)} \\ \alpha_{1k}^{(0)}\sigma_{k1}^{(0)} & \alpha_{1k}^{(0)}\sigma_{k2}^{(0)} & \alpha_{1k}^{(0)}\sigma_{k3}^{(0)} & \gamma_{11}^{(0)}/T_{avg} & \gamma_{12}^{(0)}/T_{avg} & \gamma_{13}^{(0)}/T_{avg} \\ \alpha_{2k}^{(0)}\sigma_{k1}^{(0)} & \alpha_{2k}^{(0)}\sigma_{k2}^{(0)} & \alpha_{2k}^{(0)}\sigma_{k3}^{(0)} & \gamma_{21}^{(0)}/T_{avg} & \gamma_{22}^{(0)}/T_{avg} & \gamma_{23}^{(0)}/T_{avg} \\ \alpha_{3k}^{(0)}\sigma_{k1}^{(0)} & \alpha_{3k}^{(0)}\sigma_{k2}^{(0)} & \alpha_{3k}^{(0)}\sigma_{k3}^{(0)} & \gamma_{31}^{(0)}/T_{avg} & \gamma_{32}^{(0)}/T_{avg} & \gamma_{33}^{(0)}/T_{avg} \end{bmatrix}$$

With thermoelectric properties for the inclusion $\boldsymbol{\sigma}^{(1)}, \boldsymbol{\kappa}^{(1)}, \boldsymbol{\alpha}^{(1)}$

$$\boldsymbol{E_1} = \begin{bmatrix} \sigma_{11}^{(1)} & \sigma_{12}^{(1)} & \sigma_{13}^{(1)} & \sigma_{1k}^{(1)}\alpha_{k1}^{(1)} & \sigma_{1k}^{(1)}\alpha_{k2}^{(1)} & \sigma_{1k}^{(1)}\alpha_{k3}^{(1)} \\ \sigma_{21}^{(1)} & \sigma_{22}^{(1)} & \sigma_{23}^{(1)} & \sigma_{2k}^{(1)}\alpha_{k1}^{(1)} & \sigma_{2k}^{(1)}\alpha_{k2}^{(1)} & \sigma_{2k}^{(1)}\alpha_{k3}^{(1)} \\ \sigma_{31}^{(1)} & \sigma_{32}^{(1)} & \sigma_{33}^{(1)} & \sigma_{3k}^{(1)}\alpha_{k1}^{(1)} & \sigma_{3k}^{(1)}\alpha_{k2}^{(1)} & \sigma_{3k}^{(1)}\alpha_{k3}^{(1)} \\ \alpha_{1k}^{(1)}\sigma_{k1}^{(1)} & \alpha_{1k}^{(1)}\sigma_{k2}^{(1)} & \alpha_{1k}^{(1)}\sigma_{k3}^{(1)} & \gamma_{11}^{(1)}/T_{avg} & \gamma_{12}^{(1)}/T_{avg} & \gamma_{13}^{(1)}/T_{avg} \\ \alpha_{2k}^{(1)}\sigma_{k1}^{(1)} & \alpha_{2k}^{(1)}\sigma_{k2}^{(1)} & \alpha_{2k}^{(1)}\sigma_{k3}^{(1)} & \gamma_{21}^{(1)}/T_{avg} & \gamma_{22}^{(1)}/T_{avg} & \gamma_{23}^{(1)}/T_{avg} \\ \alpha_{3k}^{(1)}\sigma_{k1}^{(1)} & \alpha_{3k}^{(1)}\sigma_{k2}^{(1)} & \alpha_{3k}^{(1)}\sigma_{k3}^{(1)} & \gamma_{31}^{(1)}/T_{avg} & \gamma_{32}^{(1)}/T_{avg} & \gamma_{33}^{(1)}/T_{avg} \end{bmatrix}$$

$$\boldsymbol{S} = \begin{bmatrix} S_{1111} & S_{1112} & S_{1113} & S_{1121} & S_{1122} & S_{1123} \\ S_{1211} & S_{1212} & S_{1213} & S_{1221} & S_{1222} & S_{1223} \\ S_{1311} & S_{1312} & S_{1313} & S_{1321} & S_{1322} & S_{1323} \\ S_{2111} & S_{2112} & S_{2113} & S_{2121} & S_{2122} & S_{2123} \\ S_{2211} & S_{2212} & S_{2213} & S_{2221} & S_{2222} & S_{2223} \\ S_{2311} & S_{2312} & S_{2313} & S_{2321} & S_{2322} & S_{2323} \end{bmatrix}$$

$$\boldsymbol{\bar{\eta}} = \begin{bmatrix} \beta & 0 & 0 & 0 & 0 & 0 \\ 0 & \beta & 0 & 0 & 0 & 0 \\ 0 & 0 & \beta & 0 & 0 & 0 \\ 0 & 0 & 0 & T_{avg}\theta & 0 & 0 \\ 0 & 0 & 0 & 0 & T_{avg}\theta & 0 \\ 0 & 0 & 0 & 0 & 0 & T_{avg}\theta \end{bmatrix}$$

**Supplementary Figure 1**

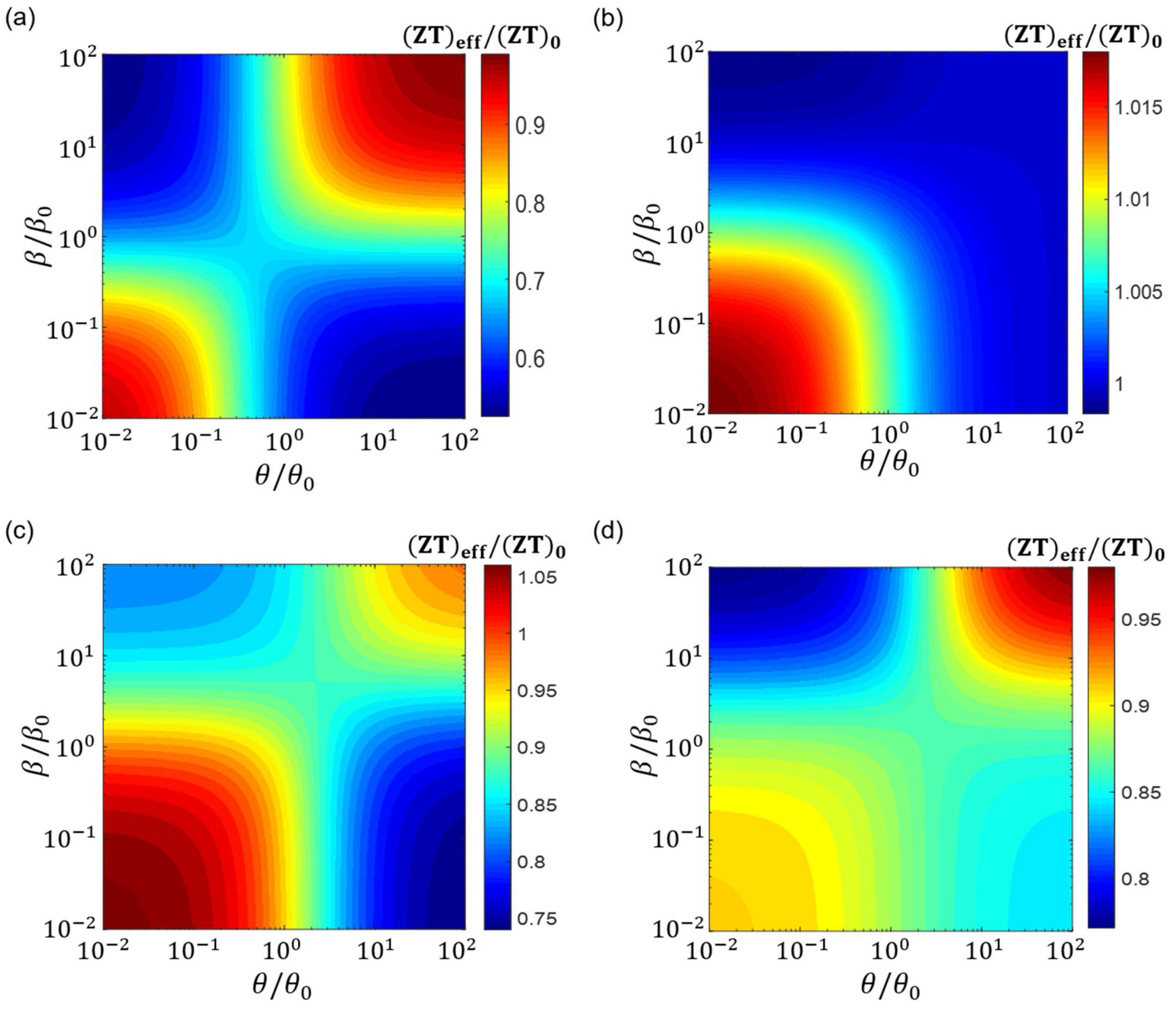

Figure 1. Normalised effective ZT at a volume fraction of 15% depending on $\beta$ and $\theta$ for (a) a $Bi_2Te_3$ matrix with a copper inclusion at 300 K, (b) a copper matrix with a $Bi_2Te_3$ inclusion at 300 K, (c) a $CoSb_3$ matrix with a $Yb-CoSb_3$ inclusion at 800 K, and (d) a $Yb-CoSb_3$ matrix with a $CoSb_3$ inclusion at 800 K (see Supplementary Table 1 for the material properties of $CoSb_3$ and $Yb-CoSb_3$).

**Supplementary Table 1**

| | $CoSb_3$[16] | $Yb-CoSb_3$[17] |
|---|---|---|
| Thermal conductivity $(\kappa)$ [W/mK] | 4 | 3.2 |
| Electrical conductivity $(\sigma)$ [S/m] | $8 \times 10^4$ | $16 \times 10^4$ |
| Seebeck coefficient $(\alpha)$ [μV/K] | 240 | 200 |
| ZT for 800 K | 0.9216 | 1.6 |

Table 1. Material properties of $CoSb_3$ and $Yb-CoSb_3$ at 800 K.